\documentclass[12pt,preprint]{aastex}

\shorttitle{Galaxy groups near PG 1115+080 and B1422+231}
\shortauthors{Grant et al.}

\begin{document}

\title{Detection of x-rays from galaxy groups associated with the gravitationally lensed systems PG 1115+080 and B1422+231}

\author{C. E. Grant, M. W. Bautz}
\affil{Center for Space Research, Massachusetts Institute of Technology, Cambridge, MA 02139}
\email{cgrant@space.mit.edu, mwb@space.mit.edu}

\and

\author{G. Chartas, G. P. Garmire}
\affil{Astronomy \& Astrophysics Department, Pennsylvania State University, University Park, PA 16802}
\email{chartas@astro.psu.edu, garmire@astro.psu.edu}

\begin{abstract}
Gravitational lenses that produce multiple images of background quasars can be an invaluable cosmological tool. Deriving cosmological parameters, however, requires modeling the potential of the lens itself.  It has been estimated that up to a quarter of lensing galaxies are associated with a group or cluster which perturbs the gravitational potential.  Detection of X-ray emission from the group or cluster can be used to better model the lens.  We report on the first detection in X-rays of the group associated with the lensing system PG~1115+080 and the first X-ray image of the group associated with the system B1422+231.  We find a temperature and rest-frame luminosity of $0.8_{-0.1}^{+0.1}$ keV and $7_{-2}^{+2} \times 10^{42}$ ergs s$^{-1}$ for PG~1115+080 and $1.0_{-0.3}^{+\infty}$ keV and $8^{+3}_{-3} \times 10^{42}$ ergs s$^{-1}$ for B1422+231.  We compare the spatial and spectral characteristics of the X-ray emission to the properties of the group galaxies, to lens models, and to the general properties of groups at lower redshift.

\end{abstract}


\section{Introduction}

Gravitational lenses that produce multiple images of background quasars can be an invaluable tool for measuring cosmological parameters, for better study of the magnified distant quasars, and for exploring the structure of the lensing galaxies.  Models of some gravitationally lensed systems require significant external shear in addition to the intrinsic asymmetry of the lensing galaxy to reproduce positions and flux ratios of the lensed images \citep*{kks97}. \citet*{kcz00} predict that a quarter of lensing galaxies are associated with a group or cluster which would perturb the gravitational potential.  Constraining the distribution of diffuse gas and dark matter in the vicinity of gravitational lenses will help remove one component of uncertainty in determining both the distance scale and the lensing galaxy properties.

Groups of galaxies may make up a significant fraction of the baryonic mass in the universe \citep*{baryons}.  Groups also occupy an interesting position in the size hierarchy between galaxies and clusters where the effects of non-gravitational heating and cooling mechanisms are comparable to gravitational effects \citep{babul02,borgani02}.  Understanding of how groups form and evolve may provide significant clues into the development and structure of both clusters and galaxies.  Detection of galaxy groups is difficult, because of the smaller number of resident galaxies and fainter group X-ray emission, especially at higher redshifts.  The greater sensitivity and improved angular resolution of {\it Chandra} and {\it XMM-Newton} are well suited to detection and study of these important objects.

We are searching for X-ray emission from groups in the fields of strong gravitational lenses using {\it Chandra's} resolving power to separate the quasar images from diffuse emission.  There are five spectroscopically confirmed galaxy group-lens associations, a few additional systems which seem likely candidates, and an ongoing program to search for these groups in the optical \citep[e.g.][]{fassnacht}.  We initially chose to concentrate on two systems, PG~1115+080 and B1422+231, as interesting candidate objects
based on their long exposure time and lensing geometry which requires external shear.  Time delays have been measured for both of these systems and the derived value of $H_0$ is dependent on the position and mass of the nearby galaxy groups which have optical properties that are favorable for detecting X-ray emission.  Both systems have been observed by previous X-ray missions \citep{chartas00}, however the strong X-ray emission from the lensed quasar and the proximity of the group to the lens made detection of the faint, diffuse group emission impossible.

PG~1115+080 is a quadruply-imaged, radio-quiet quasar at $z$ = 1.72 discovered by \citet{pgdisc}.  The lensing
galaxy has been shown to be part of a nearby group of galaxies \citep{pggrp} at $z$ = 0.310 \citep{pgz1,tonryz}.   The group properties, such as the line-of-sight velocity dispersion of $243^{+246}_{-84}$ km s$^{-1}$ and harmonic radius of $60 \pm 10 h^{-1}$ kpc \citep{fassnacht}, are generally consistent with a Hickson compact group \citep{hcg}.
Both long \citep{pgtime} and short \citep{chartas1115} time delays have been measured between the lensed images, but determination of $H_0$ is difficult because of, among other factors, the uncertainty in the location and mass profile of the associated group \citep{kk97}.  

B1422+231 is a quadruply-imaged, radio-loud quasar at $z$ = 3.62 \citep{bdisc}.
The lens is a nearby galaxy group \citep{bgrp1,bgrp2} at $z$ = 0.338 \citep{bz1,tonryz}.  The line-of-sight velocity dispersion, $535^{+314}_{-121}$ km s$^{-1}$, is at the high end for a group, while the harmonic radius, $50 \pm 10 h^{-1}$ kpc, is very consistent \citep{fassnacht}.  Time delays were measured in the radio by \citet{btime}.  \citet{hri1422} found no evidence for extended emission in a {\it ROSAT} HRI observation.  \citet*{cha1422} recently reported detection of soft X-ray emission from the group using the first of the two {\it Chandra} observations analyzed here.

We describe the {\it Chandra} observations and data reduction in \S~\ref{sec-obs}.  Detection of the group emission and analysis of the spatial characteristics are presented in \S~\ref{sec-im}.  Spectral analysis of the X-ray emission is described in \S~\ref{sec-spec}.  In \S~\ref{sec-disc} we compare our results to expectations from the optical data and the lensing models, and discuss future prospects for this type of study.  A cosmology of $H_0 = 50$ km s$^{-1}$ Mpc$^{-1}$, $\Omega_0 = 1$, and $\Lambda = 0$ is always assumed. 

\section{Observations and Analysis \label{sec-obs}}

Both objects were observed using the {\it Advanced CCD Imaging Spectrometer} \citep[ACIS;][]{acis} on the {\it Chandra X-ray Observatory} \citep{chandra}.  The target
object was placed on the ACIS-S3 back-illuminated CCD and the data were taken in the standard faint timed-exposure mode.  Table~\ref{tbl-obsdat}
lists the {\it Chandra} observational parameters.  The CCD frame time was the standard 3.2 seconds for all but the second observation of B1422+231, ObsID 1631, when the frame time was reduced to 0.8 seconds to minimize event pile up in the bright quasar images.  The estimated pile up fraction for each lensed image is 1\%--4\%, except for the earlier B1422+231 observation (ObsID 367) when the pile up fraction came close to 10\%.  Outside of the central few pixels in the quasar images, pile up is completely negligible since the expected galaxy group emission should be at least an order of magnitude fainter and diffuse.

The data were reduced using the standard {\it Chandra} tool package, CIAO, including filtering by event grade and status to remove likely background and cosmic ray afterglow events.  Times of high particle background rates were identified by examining light curves in the 0.3 -- 7 keV band.  During approximately 30\% of the exposure for ObsID 367 of B1422+231, the rate was unacceptably high (more than twice the quiescent rate) and the affected interval was removed.  Exposure times listed in Table~\ref{tbl-obsdat} were determined after this step.
Multiple observations of the same source
were merged into a single event file.  To maximize the diffuse
signal and minimize the background, the event lists were filtered
to include only photon energies between 0.5 and 2 keV, where the
relative contribution of the particle background for the ACIS-S3 detector is at its lowest.

\subsection{Image Analysis \label{sec-im}}

Two CIAO source detection algorithms were used, \verb+wavdetect+ and \verb+vtpdetect+, for a number of size scales and significance levels, but no extended sources were found.  The expected emission from the group is very weak however, so the lack of detection by generalized algorithms is not surprising.  In addition, no significant X-ray emission was detected from the individual group galaxies.  At the redshift of these groups a galaxy with a luminosity of $10^{41}$ - $10^{42}$ ergs s$^{-1}$ would produce few to no counts in our observations.

Disentangling the weak group emission from the bright quasar images is a complex task.  Even with {\it Chandra's} exquisite resolving power, the wings of the point spread function are a substantial contaminant which must be quantified and minimized.  To better study the spatial distribution of the quasar emission, a model was constructed for the lens and fit to the image using the {\it Chandra} software package Sherpa.  The lensed images were fixed to the relative positions found in the literature \citep{impey,pat99} and were represented as narrow Gaussians convolved with the {\it Chandra} point spread function (PSF).  The additional broadening introduced by the Gaussians is intended to account for errors in the aspect solution, offsets between the merged data sets, and as a gross approximation to the small distortion of the PSF caused by event pile up.  This lens model was subtracted from the original image.  The images were then smoothed with a 30 pixel FWHM Gaussian and normalized for exposure variations and instrumental features (30 pixels $\approx 15{\arcsec} \approx 42 h^{-1}$ kpc at the redshift of the groups).

\begin{figure}
\plotone{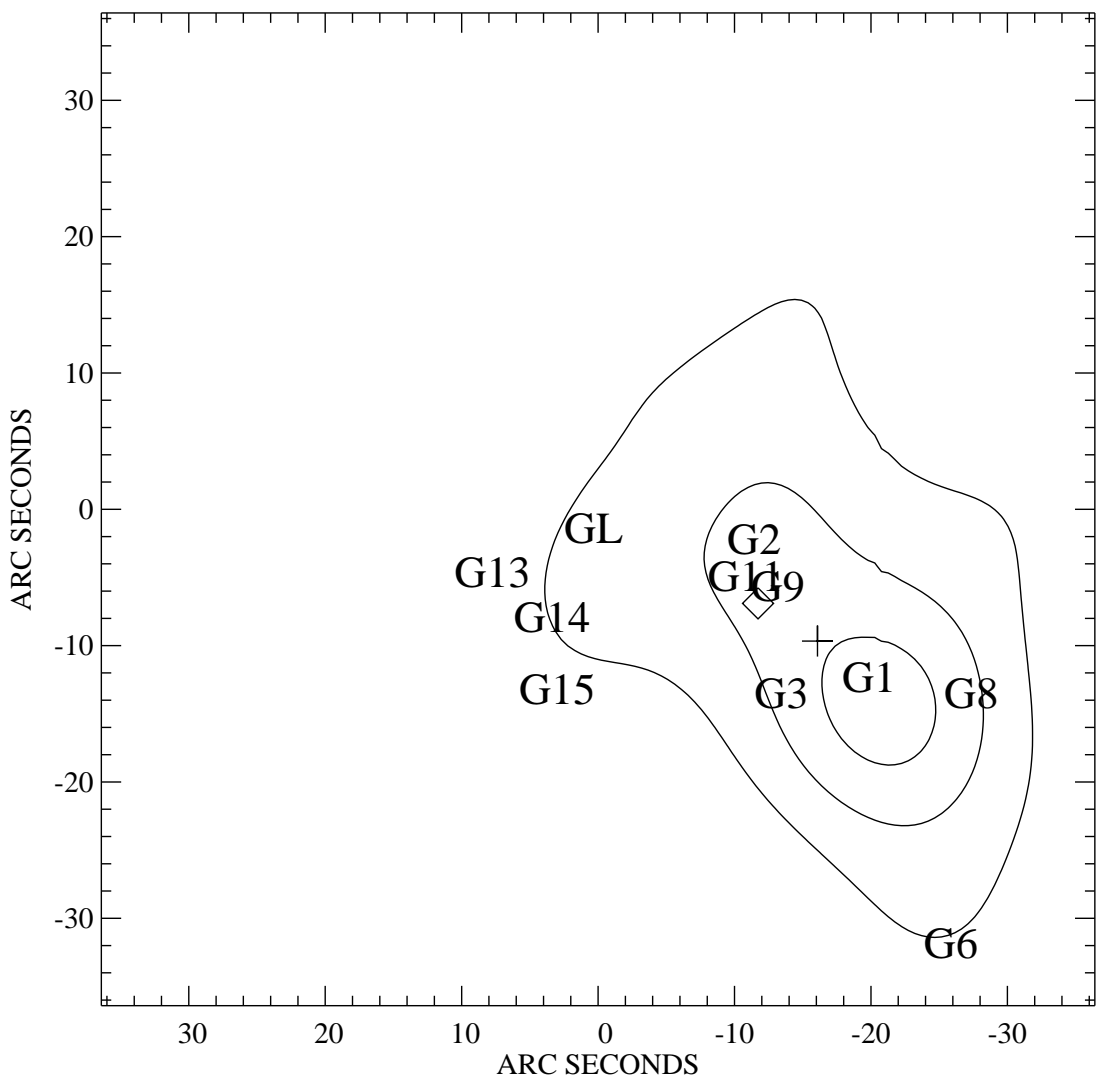}
\caption{Contours of the 0.5 - 2 keV smoothed point-source-free X-ray image in the PG~1115+080 field.  The lowest contour level is set to 2-$\sigma$ above the mean level in the field; the contours are also spaced by 2-$\sigma$.  The text labels indicate the positions of the group galaxies.  The plus sign is the flux-weighted centroid of all the galaxies as described in the text, while the diamond indicates the flux-weighted centroid of the brightest four galaxies.  ``GL'' is the position of the lensing galaxy.  Positions are relative to image C of the lens.  North is up and east is to the left.  \label{fig-1115im}}
\end{figure}

Contours of the smoothed images are shown in Figures~\ref{fig-1115im} and \ref{fig-1422im}.  The spatial scale of these figures is much larger than that of the lensed images which have a maximum separation of $1.5-2{\arcsec}$.  The positions of the known group galaxies and the lensing galaxy (``GL'') are also shown.  The galaxy positions and designations are from \citet{impey} for PG~1115+080 and from \citet{bz1} for B1422+231.  In both cases there is significant X-ray emission that appears to be extended and centered on the galaxy group.  The poor photon statistics do not allow for more complicated modeling of the source emission profile.

The PG~1115+080 group emission appears elongated with the strongest emission close to the position of the brightest group galaxy, G1, and a ridge of emission extending toward the third brightest galaxy, G2.  (The second brightest group galaxy is the lensing galaxy itself.)
The maximum X-ray emission for the PG~1115+080 group is near $\Delta\alpha = -21\arcsec$ and $\Delta\delta = -14\arcsec$ relative to image C of the gravitational lens.  The flux-weighted centroid of all the group galaxies is shown in Figure~\ref{fig-1115im} and is approximately half way between the X-ray emission peak and the clump of galaxies at the end of the extended ridge \citep{impey}.  
Also shown in Figure~\ref{fig-1115im} is the flux-weighted centroid of the four brightest galaxies which has been used to model the group in previous lens models \citep[C4 in][]{kk97} and lies near the end of the extended ridge.
Given the low statistics in the X-ray image, the X-ray centroid is reasonably consistent with both the optical centroids and the brightest group galaxy, in agreement with global properties of X-ray groups \citep{zm98,helpon2}.

The B1422+231 group emission appears more regular and compact.  The strongest emission is near the brightest group galaxy G3 and the flux-weighted centroid of the six group galaxies.
The maximum X-ray emission for the B1422+231 group is at about $\Delta\alpha = 7\arcsec$ and $\Delta\delta = -9\arcsec$ relative to image B of the gravitational lens.  The additional weaker emission to the northwest may be an extension of the group that, with higher statistics, would be connected to the brighter portion.
However, it is only 2-sigma above the background fluctuations, so it may also be a noise artifact.
The group emission is too faint to have been detected in the previous {\it ROSAT} HRI observation \citep{hri1422}.

\subsection{Spectral Analysis \label{sec-spec}}

\begin{figure}
\plotone{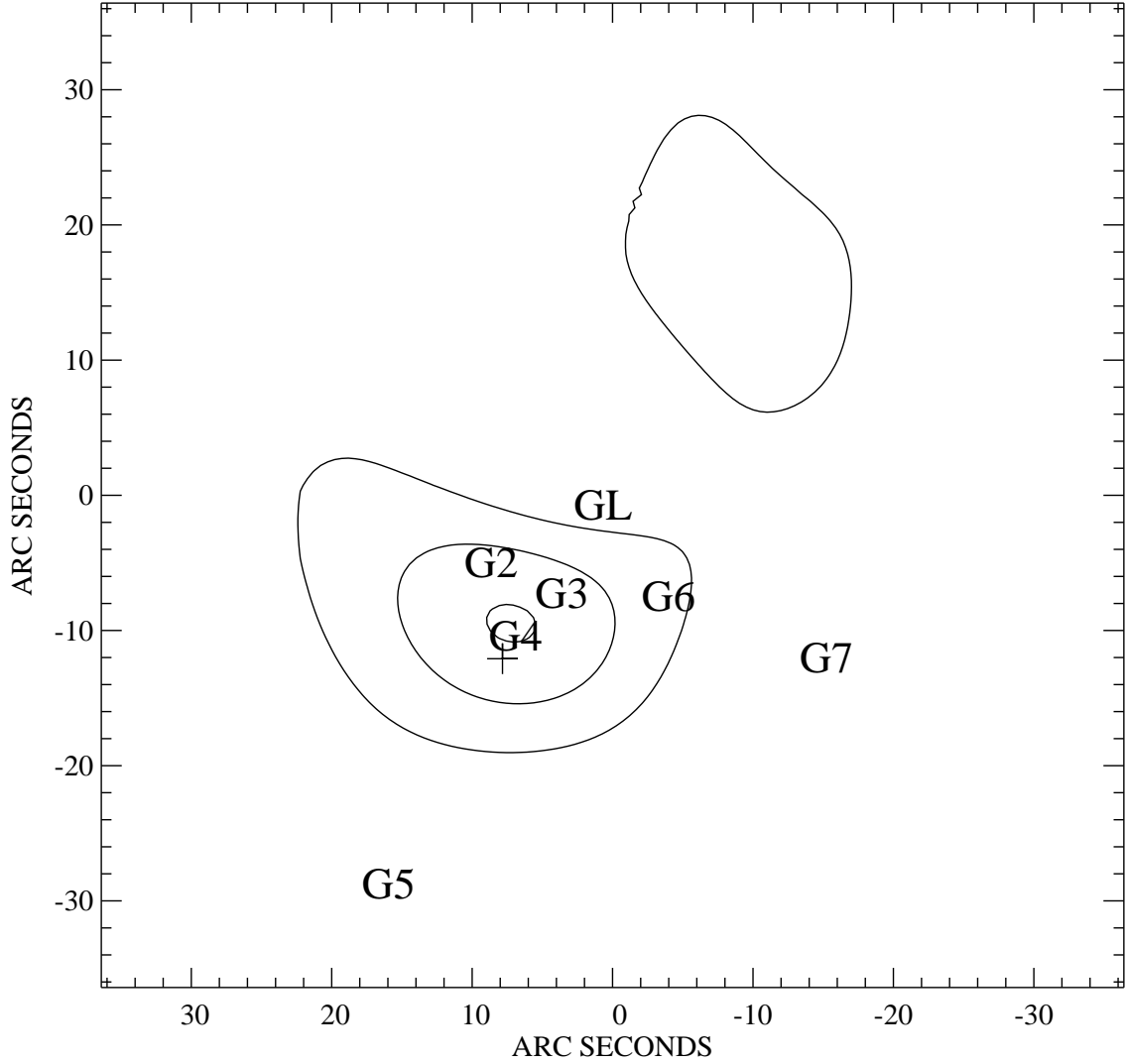}
\caption{Contours of the 0.5 - 2 keV smoothed point-source-free X-ray image in the B1422+231 field.  Positions are relative to image B of the lens.   North is up and east is to the left.  \label{fig-1422im}}
\end{figure}

Photons were extracted from a polygon region which approximates the 2-$\sigma$ contour level shown in Figures~\ref{fig-1115im} and \ref{fig-1422im}.  Photons in a circular region around the lensed quasar images were excluded.  The radius of the quasar masking region was determined by minimizing the contribution of the quasar to the extracted group spectrum to 5 - 6 counts in the 0.5 - 2 keV band.  This required a masked radius of 7 - 8{\arcsec}.
 The CIAO script \verb+acisspec+ was used to extract the photons and create the instrument response products.  The additional absorption from the time-dependent contamination of the ACIS detector was included in the effective area calculation using ACISABS.\footnote{ACISABS is available from http://www.astro.psu.edu/users/chartas/xcontdir/xcont.html.}  We find a total of 46 net source counts above a background of 46 counts for PG~1115+080 and 51 source counts above a background of 30 counts for B1422+231 in the 0.5 - 2 keV energy band. 

 We modeled the spectrum of the galaxy group as an absorbed Raymond-Smith thermal plasma with a metal abundance of 0.3 $Z_{\sun}$.  Both the Galactic $N_H$ and $z$ were set to their known values.  An additional model component was included to account for residual quasar emission with the spectral forms found in the literature \citep{gal1115,cha1422}.  The normalization of the quasar emission was adjusted to produce the number of counts predicted by the model lens image.  Because of calibration uncertainties at low energies and lack of photons at high energies, the spectral fitting was done over the energy range 0.3 - 8 keV.  The results of the spectral fitting are shown in Table~\ref{tbl-specfits}.  The fits are formally good, but the statistics are poor, particularly in the case of B1422+231.  The flux and luminosity have been adjusted to account for the expected group emission in the masked quasar region.  Because of the flat surface brightness profiles of groups, the measured luminosity in the detection region is most likely an underestimate of the true luminosity \citep{helpon2}.

Our results are in general agreement with the findings of \citet{cha1422} who found a temperature for the B1422+231 group of 0.71~keV.  They do not, however, analyze the extended emission $\sim 10{\arcsec}$ from the lens, but instead extract photons from a 3{\arcsec} region around the lensed images where quasar contamination is orders of magnitude worse.  Their lens X-ray contours seem more extended in the NW direction, which may argue for the reality of the larger northern feature seen in Figure~\ref{fig-1422im}.  In addition, since they do not mention any correction for event pile up yet extract the quasar photons from very small regions around the image cores, we must assume that their fitted power law index for the quasar spectrum, which we use in our spectral analysis, is too hard.  Because of the low photon statistics for the group emission however, this does not have a significant influence on our derived group properties.

\section{Discussion \label{sec-disc}}

We have detected X-ray emission from the two groups associated with the gravitational lenses.  We can now compare the observed X-ray luminosity and temperature for the groups to the optical group characteristics, to global properties of groups at low redshift, and to lens models in the literature.  We use the $\sigma-T_X$, $\sigma-L_{bol}$, and $L_{bol}-T_X$ relations from \citet{helpon}, who fit a large sample of compact and loose groups.  Other scaling relations \citep[e.g.][]{girscal,xuewu,mgscal} yield comparable results. 

\begin{figure}
\plotone{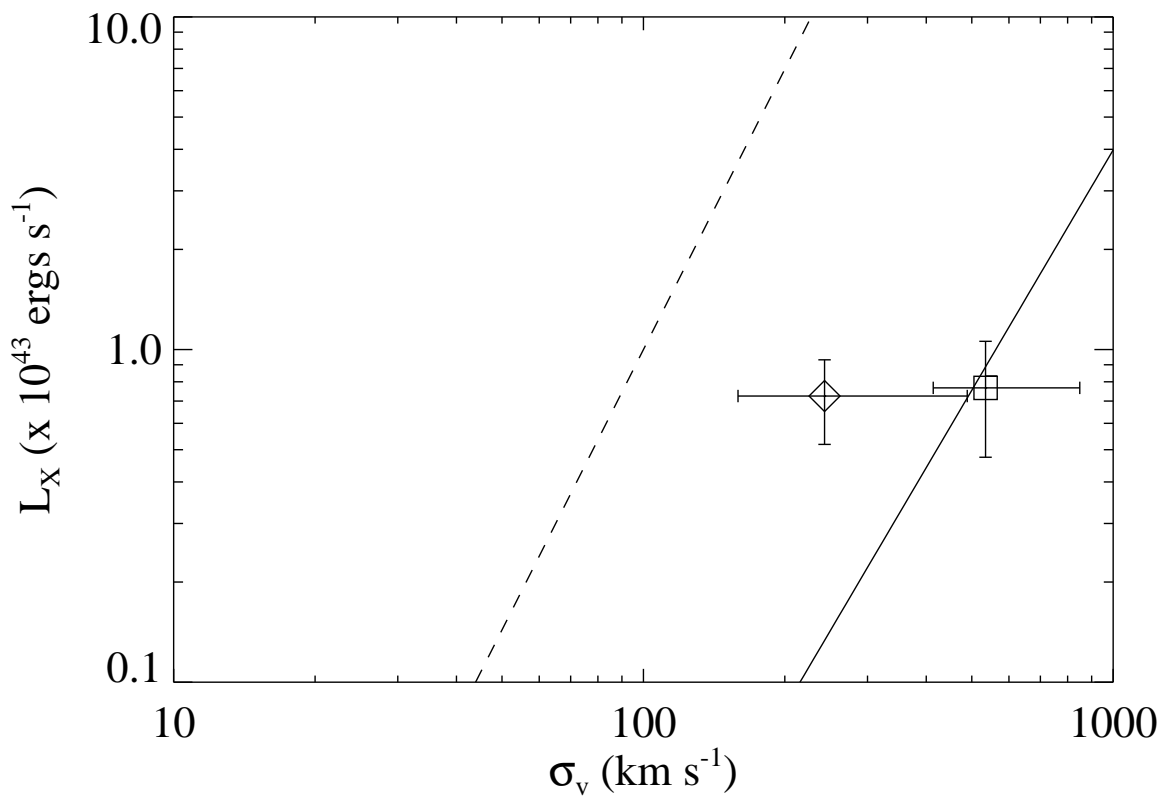}
\caption{Comparison of the velocity dispersion and luminosity of the two galaxy groups to the scaling relation of \citet{helpon}.  PG~1115+080 is indicated by a diamond symbol and B1422+231 by a square symbol.  The dotted line indicates the upper 1-$\sigma$ error bounds of the fitted relation.  The lower 1-$\sigma$ error bounds of the fitted relation are off the lower-right edge of the plot.  The data error bars are 68\% confidence.  \label{fig-sl}}
\end{figure}

\begin{figure}
\plotone{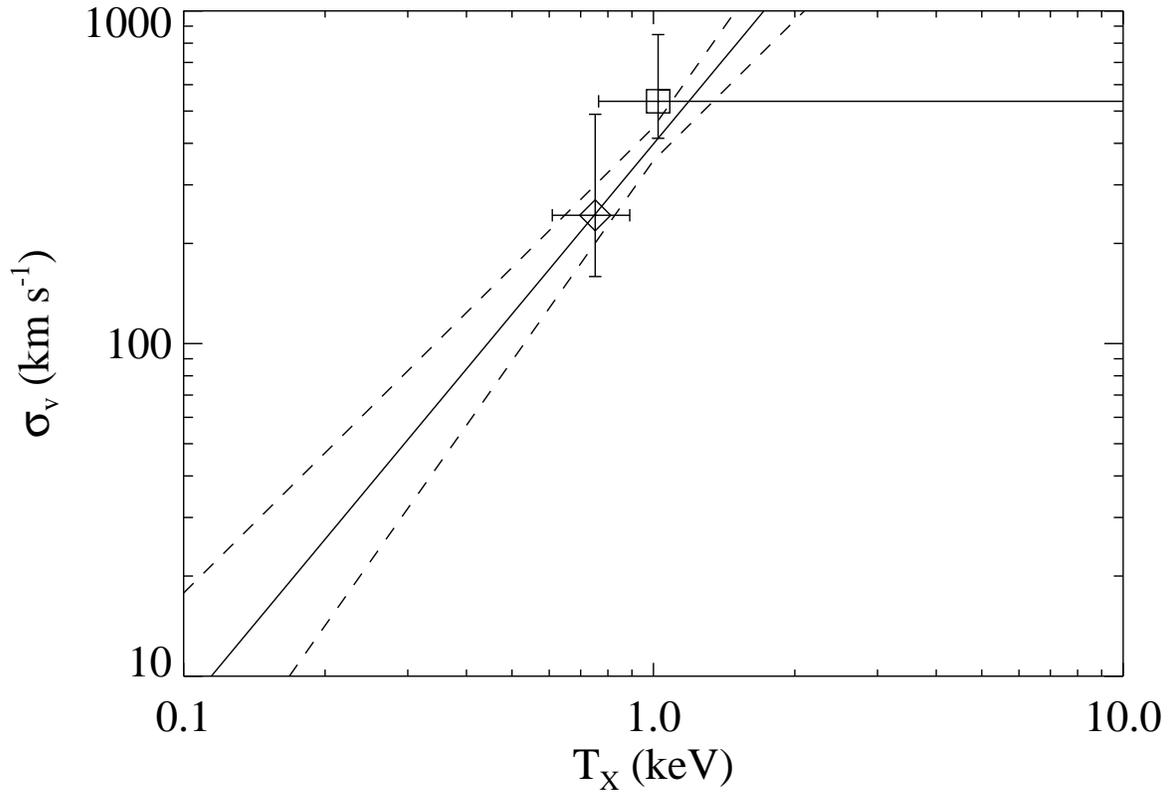}
\caption{Comparison of the temperature and velocity dispersion of the two galaxy groups.  Symbols are the same as Figure~\ref{fig-sl}. \label{fig-st}}
\end{figure}

\begin{figure}
\plotone{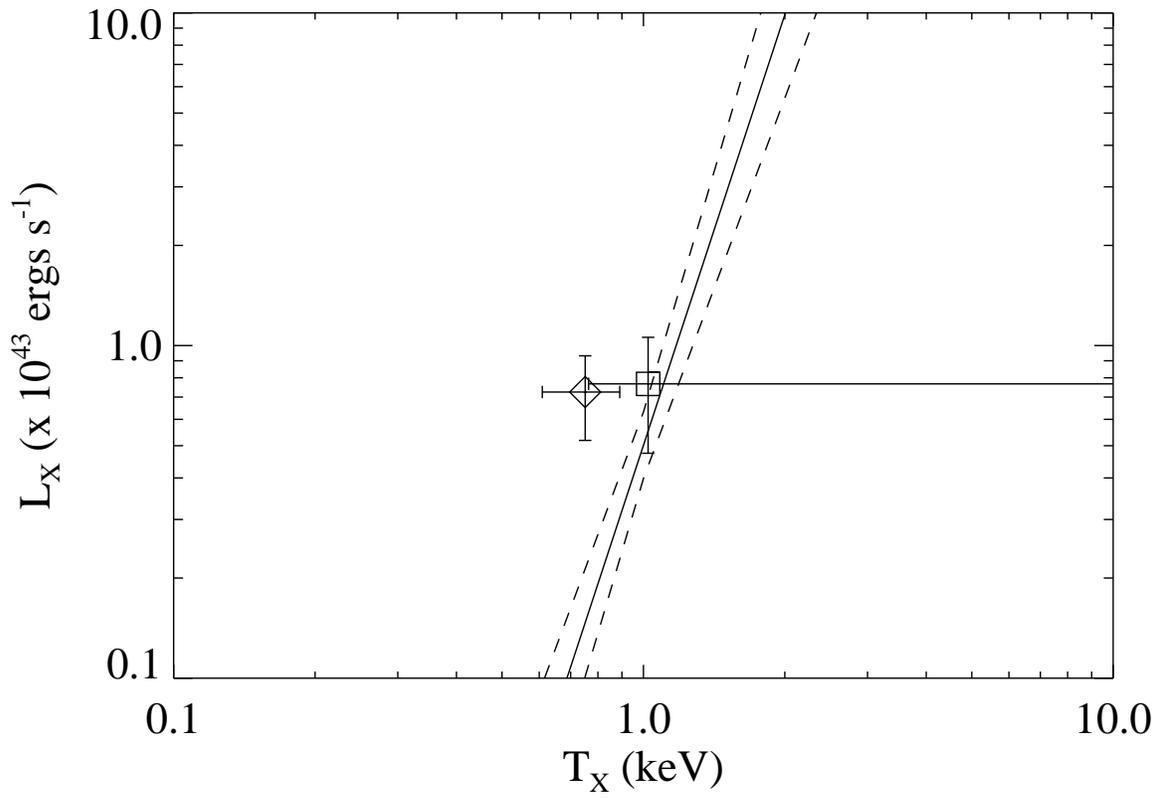}
\caption{X-ray luminosity and temperature of the galaxy groups.  Symbols are the same as Figure~\ref{fig-sl}. \label{fig-lt}}
\end{figure}

Figures~\ref{fig-sl}, \ref{fig-st}, and \ref{fig-lt} compare the group X-ray luminosity, temperature, and the galaxy velocity dispersion to the scaling relations for low redshift groups from \citet{helpon}.  Both groups are consistent with the scaling relations within the errors of the data and the fit.  The temperature for B1422+231 is poorly constrained, but the best fit value of 1~keV is consistent with that expected from a normal group.  PG~1115+080, while formally consistent, is somewhat overluminous given the measured velocity dispersion and temperature.  These results are in general agreement with \citet{jones02}, who found no evidence of group evolution out to $z$ of 0.5.

Given a model for the group potential, the velocity dispersion of the group, $\sigma_v$, and its position can be used to predict $\gamma$, the external shear at the lens from the group.  The same group scaling relations can then further predict the shear given the X-ray temperature or luminosity.  For a singular isothermal sphere (SIS) model, $$\gamma = \frac{2\pi}{r} \frac{D_{ls}}{D_{os}}\left(\frac{\sigma_v}{c}\right)^2 $$ where $r$ is the angular distance from the group to the main lensing galaxy, and $D_{ls}$ and $D_{os}$ are the angular diameter distances lens-source and observer-source.

Gravitational lens models of the PG~1115+080 system require an external shear of order $\gamma \sim 0.1$ in the general direction of the galaxy group center to fit the observed image positions \citep[e.g.][]{kks97}.  Modeling the group as an SIS located at the position of maximum X-ray emission, the shear at the lensing galaxy from the group potential is of order $\gamma \sim 0.01$ using the measured velocity dispersion or using the $\sigma-T_X$ scaling relation and the measured temperature; the shear is of order $\gamma \sim 0.1$ using the $\sigma-L_{bol}$ scaling relation and the measured luminosity.  
Moving the group center closer to the lensing galaxy would increase the predicted value of gamma.
In order to produce an external shear of order $\gamma \sim 0.1$ given the measured velocity dispersion and X-ray temperature, the group centroid must be no further than $14\arcsec$ from the lensing galaxy.
When \citet{kk97} allowed the group position to be a free parameter in their lens models, the best-fit model had $(d,\theta) = (25\farcs2, -125\arcdeg)$, very close to the X-ray emission peak.
Using better image and lens positions, \citet{impey} found a group position of $(10\arcsec, -113\arcdeg)$, consistent with C4, the luminosity centroid of the brightest four group galaxies.
The X-ray group emission profile appears elongated, so the group potential may be better represented by an elliptical rather than spherical model.
The external shear can be derived for an elliptical potential, such as that of \citet{bk87}.
The magnitude of the shear is then also dependent on the ellipticity of the group potential and the angle between the group major axis and the lens direction.
If, as may be true in this case, the major axis is aligned along the direction to the lens, the predicted shear increases.
A more accurate model for the group potential would include both the apparent ellipticity as well as a more refined X-ray centroid.
The predicted shear from this model would likely be in agreement with the estimates from lensing models.

The lensed images in B1422+231 are highly asymmetric and so require substantial external shear of order $\gamma \sim 0.2$ in the direction of the galaxy group \citep[e.g.][]{kks97}.  Assuming an SIS group at the position of maximum X-ray emission with the measured properties, the external shear from the group potential at the lensing galaxy is in general agreement with lens models.

In both cases, the group parameters derived from the optical properties of the galaxies seem to be in general agreement with those from the X-ray properties, however there are interesting problems remaining.  The elongated structure of the PG~1115+080 group, if real, may be important is determining how much shear and convergence the group contributes to the lens.  Better counting statistics would certainly help constrain the group X-ray parameters, and because of the small angular distance and large dynamic range between the quasar images and the group, the high spatial resolution of {\it Chandra} will be required in future investigations.

\acknowledgments

We thank the anonymous referee for constructive and helpful comments.  We have made use of the CfA-Arizona Space Telescope Lens Survey (CASTLES) website (http://cfa-www.harvard.edu/castles) maintained by C. S. Kochanek, E. E. Falco, C. Impey, B. McLeod, H.-W. Rix.  This work was supported by NASA contracts NAS8-37716 and NAS8-38252.

\clearpage

\begin{deluxetable}{lrcr}
\tablecaption{Observation Data \label{tbl-obsdat}}
\tablehead{
\colhead{Name} & \colhead{Observation ID} & \colhead{Observation Date} & 
\colhead{Exposure Time (s)}
}
\startdata
PG 1115+080 &363  &2000 Jun 02 &26492 \\
            &1630 &2000 Nov 03 &9825  \\
B1422+231   &367  &2000 Jun 01 &17864 \\
            &1631 &2001 May 21 &10651 \\
\enddata
\end{deluxetable}

\clearpage

\begin{deluxetable}{lcc}
\tablecaption{Group X-ray Emission Properties \label{tbl-specfits}}
\tablehead{ & \colhead{PG 1115+080} & \colhead{B1422+231}}
\startdata
Group redshift &0.310 &0.338 \\
Galactic $N_{H}$ ($10^{20}$ cm$^{-2}$) &4.0 &2.7 \\
Group (Background) 0.5 - 2 keV Counts      &46 (46) &51 (30) \\
$kT$ (keV)                     &$0.8_{-0.1}^{+0.1}$ &$1.0_{-0.3}^{+\infty}$\\
Absorbed 0.5 - 2 keV flux ($10^{-15}$ ergs cm$^{-2}$ s$^{-1}$) &$4.1_{-1.2}^{+1.2}$ &$3.2_{-1.2}^{+1.2}$ \\
Rest-frame bolometric luminosity (10$^{42}$ ergs s$^{-1}$) &$7.2_{-2.1}^{+2.1}$ &$7.7_{-2.9}^{+2.9}$\\ 
X-ray emission peak $(d, \theta)$ &($25\arcsec, -123\arcdeg)$ &$(11\arcsec, 144\arcdeg)$ \\
\enddata
\tablecomments{Errors are 1-$\sigma$ (68\%) confidence level.  Redshift and Galactic column density are fixed at their known values.  The position is relative to image C for PG~1115+080 and image B for B1422+231.  The position angle is measured north through east.  $H_0$ = 50 km s$^{-1}$ Mpc$^{-1}$ and $\Omega_0$ = 1.0}
\end{deluxetable}

\clearpage

\clearpage

\clearpage

\clearpage

\clearpage

\end{document}